\documentclass{aastex63}
\usepackage{graphicx}
\usepackage{lineno} 
\usepackage{subfigure}
\usepackage{multirow}
\usepackage{mathptmx} 
\usepackage{hyperref} 
\usepackage{chngcntr} 

\hypersetup{linkcolor=blue,citecolor=blue,filecolor=cyan,urlcolor=blue}

\newcommand{\mytitle}{Core mass function of a single giant molecular cloud complex with $\sim10^4$ cores}

\newcommand{\NJU}{School of Astronomy and Space Science, Nanjing University, 163 Xianlin Avenue, Nanjing 210023, People's Republic of China}
\newcommand{\NJUMath}{Department of Mathematics, Nanjing University, 163 Xianlin Avenue, Nanjing 210023, People's Republic of China}
\newcommand{\NJULab}{Key Laboratory of Modern Astronomy and Astrophysics (Nanjing University), Ministry of Education, Nanjing 210023, People's Republic of China}
\newcommand{\CfA}{Center for Astrophysics $\vert$ Harvard \& Smithsonian, 60 Garden Street, MS 42, Cambridge, MA 02138, USA}

\newcommand{\hii}{H{\scriptsize\ II}}

\newcommand{\msun}{$M_{\odot}$} 
\newcommand{\lsun}{$L_{\odot}$}

\newcommand{\beamColden}{20\arcsec}
\newcommand{\beamT}{36\arcsec}

\newcommand{\NumClumpGsOrig}{12,427}
\newcommand{\NumClumpGs}{8,431}
\newcommand{\NumClumpCfOrig}{4,974} 
\newcommand{\NumClumpCf}{4,479} 
\newcommand{\NumClumpGsCore}{48} 

\newcommand{\ClumpGsSizeMin}{0.14} 
\newcommand{\ClumpGsSizeMax}{0.57}
\newcommand{\ClumpGsMassMin}{0.1} 
\newcommand{\ClumpGsMassMax}{1300}

\newcommand{\mTransitGs}{10} 
\newcommand{\slopeGs}{2.30}
\newcommand{\slopeErrorGs}{0.04}

\newcommand{\ClumpGsMassNighty}{100}
\newcommand{\ClumpGsMassEighty}{20}

\newcommand{\NumCoreRaw}{923} 
\newcommand{\NumCore}{200} 
\newcommand{\NumCoreIR}{51} 
\newcommand{\NumCoreClump}{180} 

\newcommand{\CoreSizeMin}{0.008} 

\newcommand{\CoreSizeMax}{0.05}
\newcommand{\CoreFluxMin}{0.006} 

\newcommand{\CoreFluxMax}{1.2}
\newcommand{\CoreLIRMin}{0.01} 

\newcommand{\CoreLIRMax}{1160}
\newcommand{\CoreTMin}{14} 
\newcommand{\CoreTMed}{21}
\newcommand{\CoreTMax}{58}
\newcommand{\CoreMassMin}{0.2} 
\newcommand{\CoreMassMed}{2.2}
\newcommand{\CoreMassMax}{84}

\newcommand{\indexClumpMassCoreMass}{0.27}
\newcommand{\indexClumpMassCoreMassErr}{0.14}
\newcommand{\rValue}{0.27}

\newcommand{\indexClumpMassCoreMassMC}{1.00}
\newcommand{\indexClumpMassCoreMassMCErr}{0.11}

\newcommand{\pValue}{0.014} 

\begin{document}

\title{\mytitle}

\author[0000-0002-6368-7570]{Yue Cao}
\affiliation{\NJU}\affiliation{\NJULab}\affiliation{\CfA}
\author[0000-0002-5093-5088]{Keping Qiu$^{\ast}$}
\affiliation{\NJU}\affiliation{\NJULab}
\author[0000-0003-2384-6589]{Qizhou Zhang}
\affiliation{\CfA}
\author[0000-0001-6630-0944]{Yuwei Wang}
\affiliation{\NJU}\affiliation{\NJULab}
\author[0000-0002-3558-4523]{Yuanming Xiao}
\affiliation{\NJUMath}

\correspondingauthor{Keping Qiu}
\email{kpqiu@nju.edu.cn}


\begin{abstract} 

Similarity in shape between the initial mass function (IMF) and the core mass functions (CMFs) in star-forming regions prompts the idea that the IMF originates from the CMF through a self-similar core-to-star mass mapping process. To accurately determine the shape of the CMF, we create a sample of \NumClumpGs\ cores with the dust continuum maps of the Cygnus X giant molecular cloud complex, and design a procedure for deriving the CMF considering the mass uncertainty, binning uncertainty, sample incompleteness, and the statistical errors. The resultant CMF coincides well with the IMF for core masses from a few \msun\ to the highest masses of 1300 \msun\ with a power-law of ${\rm d}N/{\rm d}M\propto M^{-\slopeGs\pm\slopeErrorGs}$, but does not present an obvious flattened turnover in the low-mass range as the IMF does. More detailed inspection reveals that the slope of the CMF steepens with increasing mass. Given the numerous high-mass star-forming activities of Cygnus X, this is in stark contrast with the existing top-heavy CMFs found in high-mass star-forming clumps. We also find that the similarity between the IMF and the mass function of cloud structures is not unique at core scales, but can be seen for cloud structures of up to several pc scales. Finally, our SMA observations toward a subset of the cores do not present evidence for the self-similar mapping. The latter two results indicate that the shape of the IMF may not be directly inherited from the CMF. 
\end{abstract}


\section{Introduction}\label{sec:intro} 

Stellar initial mass function (IMF) depicts the mass distribution for a population of newborn stars and appears to be universal at least within the local galaxies \citep{2014prpl.conf...53O}, yet its origin is still not well understood \citep{2013pss5.book..115K,2014prpl.conf...53O}. Stars assemble their masses from molecular clouds. Observational studies on Galactic star-forming regions found that the mass distribution of cores---dense, elliptical structures on 0.01--0.1 pc scales\footnote{The definition of cores can vary and their sizes can span more than one order of magnitude in different studies, so specifications are needed. In this work our core sample have a median size of 0.3 pc.}---has a shape similar to the IMF but is shifted systematically towards higher masses by a factor of $\sim3$ \citep{1998A&A...336..150M,1998ApJ...508L..91T,2007A&A...462L..17A,2015A&A...584A..91K}. This discovery prompts the intuitive idea that IMF originates from the core mass function (CMF) via a self-similar core-to-star mass mapping process, i.e. the probability that a core forms a star is constant as long as the core-to-star mass ratio is constant \citep{1998A&A...336..150M,2007A&A...462L..17A,2018NatAs...2..478M}. To understand the physics that drives this resemblance and whether (and if so, how) it is related to the origin of the IMF, it is essential to enlarge the sample size and accurately determine the shape of CMF within a single giant molecular cloud complex.

The Cygnus X giant molecular cloud complex (hereafter Cyg X) is one of the nearest \citep[distance from the Sun of $\sim1.4$ kpc;][]{2012A&A...539A..79R}, largest (projected dimension of $\sim130$ pc), and most massive molecular cloud complexes in the Milky Way. It has a total molecular mass of $\sim3\times10^6$ \msun\ \citep[from][]{2019ApJS..241....1C}, and is representative of the active high-mass star-forming regions in the Galaxy by its numerous \hii\ regions \citep{1991A&A...241..551W}, OB associations \citep{2001A&A...371..675U}, and star-forming filaments and clumps \citep{2012A&A...543L...3H}, and is thus an ideal target for constructing a large sample of cloud structures. The identification of cores in Cyg X was conducted on an $\rm H_2$ column density ($N_{\rm H_2}$) map illustrated in Fig. \ref{fig:panorama}, which was generated by \citep{2019ApJS..241....1C} with the \emph{hirescoldens} procedure \citep{2012A&A...542A..81M} that fits pixel-by-pixel the \emph{Herschel} continuum images with a graybody thermal dust emission model \citep{1983QJRAS..24..267H}. The map provides abundant and detailed information of the spatial distribution of the cold molecular gas in Cyg X with a dynamic range of 1,300 in spatial scale (0.14--180 pc) and 1,000 in column density ($4\times10^{20}$--$4\times10^{23} \rm cm^{-2}$), which makes it feasible to study the cloud structures within a single-distance giant molecular cloud with a wide range of physical properties and high statistical significance.

\section{Analysis}\label{sec:analysis}

We design a non-parameterized procedure shown in Fig. \ref{fig:flowchart} dedicated to derive a ``true'' probability distribution of source masses and its uncertainties simultaneously from a source sample. The procedure consists of the following steps: (1) extract sources on a column density map to obtain a real source sample; (2) assess the uncertainty of the mass estimation and derive a series of samples with randomized masses from the real sample; (3) generate a series of randomized mass bins and derive the (raw) mass frequency distributions of the randomized samples; (4) derive a series of pseudo-samples, insert and extract them on the column density map to obtain the extracted pseudo-samples; (5) derive a series of completeness levels as a function of mass by spatially matching the pseudo-samples and the extracted pseudo-samples; (6) correct the raw mass frequency distributions for completeness and obtain a series of corrected mass frequency distributions; (7) reconstruct the true mass probability distribution and its uncertainties based on the statistics of the corrected mass frequency distributions. We argue that this procedure is superior to drawing raw mass distributions (as is done in some literatures) by, first, addressing the sample incompleteness; second, properly estimating all the sources of uncertainties of the mass probability distribution in its derivation process (i.e. mass uncertainty, binning uncertainty, and the uncertainty in completeness levels addressed in steps (2), (3), and (6), respectively); third, greatly reducing the random statistical errors of the mass probability distribution (via large number of Monte-Carlo simulations; see below). Detailed implementations and results of each step are described in Sects. \ref{subsec:sample} and \ref{subsec:completeness}, and more detailed setups, results, and plots of the Monte-Carlo simulations in these steps are elaborated in Appendix \ref{app:robust}. In Sect. \ref{subsec:clumpfind} we examine the dependence of our results on the source-extraction algorithms. Readers can refer to Sect. \ref{subsec:results} for the final results of the derived CMF.

\subsection{Source extraction, mass randomization, and bins randomization}\label{subsec:sample} 

We used the \emph{getsources} algorithm \citep{2012A&A...542A..81M} to extract cores in Cyg X. Designed for identifying dense cloud structures in star-forming regions, \emph{getsources} removes the large-scale background of an image, detects sources by decomposing the image into components on single spatial scales, and measures their properties (coordinates, full-width-at-half-maximum (FWHM) diameters, peak column densities, and masses) with the 2-dimensional Gaussian model. After a blind run of \emph{getsources} with the $N_{\rm H_2}$ map as input, a total of \NumClumpGsOrig\ sources were identified. Then we apply the following criteria to select robust detections for deriving the CMF: (1) core mass $M_{\rm core}\geq$0.07 \msun; (2) relative core mass uncertainty given by \emph{getsources} $\delta M/M\le1$; (3) core FWHM diameter no larger than 0.6 pc; (4) cores should be $\geq$50\arcsec\ away from the map borders to avoid artifacts. Criteria (1) and (3) are dedicated to exclude the implausibly large/small values of the parameters derived by \emph{getsources}. With these criteria \NumClumpGs\ cores with robust masses and diameters were selected and their spatial distributions in Cyg X are shown in Fig. \ref{fig:panorama}. Statistically, these cores have FWHM diameters ranging from \ClumpGsSizeMin\ to \ClumpGsSizeMax\ pc and masses ranging from \ClumpGsMassMin\ to \ClumpGsMassMax\ \msun. Monte-Carlo simulations show that the 1-$\sigma$ uncertainty in the core mass evaluation is $\sim$0.23 dex (see Appendix \ref{subapp:mass}). 

To estimate the contribution of the mass uncertainty to the uncertainty of CMF, we generate 100 core samples with the same sample size as the real core sample (\NumClumpGs\ cores) and with randomized masses. The mass randomization is implemented by multiplying the real core masses with random factors that simulates the effects of mass uncertainty. The overall effect of this process is that the mass values are randomly shifted by $\pm0.23$ dex on average. Detailed processes and results of the mass randomization are described in Appendix \ref{subapp:mass}. We then derive 100 mass frequency distributions with the 100 randomized samples and 100 randomized mass-bin sets. The mass bins are randomized so that the contribution of mass binning to the uncertainty of the mass probability distributions can be addressed. Each randomized bin set has a bin number close to the square root of the sample size, and a series of random bin centers log-uniformly distributed within the mass range of the sample. For the convenience of further calculations, The 100 mass frequency distributions are then resampled to a fixed set of 90 bins with uniform bin widths in logarithmic mass scales, and to be followed by the completeness correction. 

\subsection{Completeness correction and the derivation of the true mass probability distribution}\label{subsec:completeness} 

The completeness level of a sample extracted from a map can never reach 100\% due to the noise, background structures on the map, and the performance of the extraction algorithm, and is primarily an increasing function of source mass/flux. Thus the mass/flux distribution of the sample should be corrected by the completeness levels to reflect the true distributions. To derive the completeness level of the core sample of Cyg X as a function of mass, $\eta(M_{\rm core})$, we insert pseudo-cores with a wide range of known masses on the $N_{\rm H_2}$ map and calculate the fractions of the detected ones by \emph{getsources}. This insertion-extraction experiment is done for 10 times to obtain a better estimate of the completeness level and to derive their statistical errors. Detailed setups and results of the experiments are described in Appendix \ref{subapp:complete}, and the completeness levels of the core sample are presented in Fig. \ref{fig:detect_rate}. As can be seen, the complete level can be best described as an S-shaped increasing function of mass. The core sample is 90\% complete for $M_{\rm core}\geq\ClumpGsMassNighty$ \msun, and 80\% complete for $M_{\rm core}\geq\ClumpGsMassEighty$ \msun. 
 
We corrected the 100 raw mass frequency distributions in Sect. \ref{subsec:sample} with the 10 completeness level functions via ${\rm d}N^{\rm c}/{\rm d}M_{\rm core}=[{\rm d}N/{\rm d}M_{\rm core}]/\eta(M_{\rm core})$. Each completeness level function is used for 10 out of the 100 mass frequency distributions so there is no shortage of completeness levels. Fig. \ref{fig:CMF_all_scale} shows the 100 corrected mass frequency distributions as gray dots. 

As a final step, the true probability distribution of the core mass and its 1-$\sigma$ upper/lower error bars are derived as the median values and the lengths of the upper/lower 34\%-percentile intervals of the 100 corrected mass frequency distributions in individual mass bins, respectively. Fig. \ref{fig:CMF_gs} shows the mass probability distribution and the error bars as the final results. It is clear that the reconstructed distribution is much smoother and steadier than the raw distribution of the core masses, as the former has much fewer statistical errors thanks to the large number (100) of randomized mass distributions in its derivation procedure. To illustrate the individual contributions of the sources of uncertainties (i.e. mass uncertainty, binning uncertainty, and the uncertainty of completeness levels) to the error bars of the mass probability distribution, we go through the processes in Fig. \ref{fig:flowchart} again but with only one type of randomization implemented each time, and plot the resultant error bars in Fig. \ref{fig:error_bar}. The completeness levels are the most dominant source of uncertainty in the low-mass range, which is reasonable since their values are close to 0 and thus the relative errors are large. In the high-mass range, the uncertainties are dominated by the mass and the binning uncertainties. It is also worth mentioning that a majority of similar studies use the Poisson counting error (proportional to $\sqrt{N}$ for $N$ counts in a bin) to estimate the binning uncertainty, which can provide similar results to what we present with simulations, but has been proven to be mathematically not rigorous \citep[e.g. the error should not be zero when the count in a bin is zero; see more in][]{2012EPJP..127...24A}. 

\subsection{Testing the results with an alternative source-extraction algorithm}\label{subsec:clumpfind}

So far we have been using \emph{getsources} to extract both real and pseudo-cores in the procedure of deriving the CMF. To examine the dependence of our results on extraction algorithms, we derive the CMF and its uncertainty following the procedure in Fig. \ref{fig:flowchart} again but with an alternative algorithm \emph{clumpfind} in the PyCupid\footnote{\url{https://github.com/ChileanVirtualObservatory/pycupid}} \citep{2007ASPC..376..425B} python package. \emph{Clumpfind} detects sources in an image or a data cube by identifying local peaks and attributing adjacent pixels to the peaks while tracing down the descending contours of the image/cube \citep{1994ApJ...428..693W}. A blind run of \emph{clumpfind} on the $N_{\rm H_2}$ map yields \NumClumpCfOrig\ extracted sources. The reason why the sources are fewer than those identified by \emph{getsources} is that \emph{clumpfind} has poorer detection performance in the low-mass range (see Fig. \ref{fig:detect_rate}). We further select \NumClumpCf\ robust cores that are $\ge50$\arcsec\ away from the map borders to avoid artifacts. These robust cores have masses ranging from 0.49 to 1553 \msun, and a median FWHM size of 0.29 pc. The resultant mass frequency distributions and the mass probability distribution of the core sample are presented in Fig. \ref{fig:CMF_all_scale}, and a direct comparison with the results of \emph{getsources} is illustrated in Fig. \ref{fig:compare}. As can be seen, the overall shape of the CMF derived with \emph{clumpfind} is very close to that derived with \emph{getsources}, except that the mass range of the \emph{clumpfind} CMF is narrower due to the poorer sensitivity. For $M_{\rm core}\gtrsim 15$ \msun, the \emph{clumpfind} CMF is higher than the \emph{getsources} CMF since \emph{clumpfind} has slightly better detection performance in this mass range (see Fig. \ref{fig:detect_rate}). The shapes of the CMFs will be further analyzed in Sect. \ref{subsec:results}.

\section{Discussion}\label{sec:discussion} 

\subsection{Shape of the CMF in Cyg X}\label{subsec:results} 

As is shown in Fig. \ref{fig:CMF_gs}, the overall shape of the CMF can be roughly approximated as a single power-law function to the first order. We fit the high-mass part of the CMF with a power-law function to determine its power-law index. The starting point of the CMF for the fitting is determined to be \mTransitGs\ \msun\ using the Ramer--Douglas--Peucker algorithm \citep{Douglas1973AlgorithmsFT}, which downsamples a curve into fewer most representative points. We use the \emph{curve\_fit} function in the SciPy package \citep{scipy} to implement the fitting considering both the values and the uncertainties of the CMF. The fitting yields ${\rm d}N/{\rm d}M_{\rm core}\propto M_{\rm core}^{-(\slopeGs\pm\slopeErrorGs)}$ for the high-mass part, and the slope index is close to the canonical values (2.3--2.35) of the IMFs \citep{1955ApJ...121..161S,2001MNRAS.322..231K}. In Fig. \ref{fig:CMF_gs} we also plot the shifted IMFs of \citet{2001MNRAS.322..231K} and \citet{2005ASSL..327...41C} with their power-law transition points pinned at the starting point of fitting of the CMF. We see that both IMFs are close to the CMF for core masses above a few \msun. In the low-mass part, the CMF deviates from the IMFs by much larger sample numbers. 

With a closer look at the CMF we can find that the slope of the CMF is steepening with increasing mass, which makes the CMF deviating from the single power-law fit (Fig. \ref{fig:CMF_gs}). This feature is robustly resolved for CMFs for the first time, thanks to the statistical significance of our core sample and the methods we used to reduce statistical errors. To further illustrate this we plot in Fig. \ref{fig:slope} the power-law indexes as a function of mass derived with local consecutive data points of the CMF, which again presents a robust increasing trend. Recent observations with submm/mm interferometers toward distant ($\ge1$ kpc) high-mass star-forming clumps reveal that the CMFs in these regions have power-law slopes shallower than that of the IMF, presenting a top-heavy feature that is different from the CMFs of nearby low-mass star-forming regions (which resemble the IMF) \citep{2018NatAs...2..478M,2019ApJ...886..102S,2020ApJ...894L..14L}. On the other hand, our finding suggests that on larger scales, the CMF of Cyg X is \emph{not} top-heavy in the high-mass range, despite the numerous high-mass star-forming regions therein. In fact, most of the cores with $M_{\rm core}\ge200$ \msun\ where the CMF shows the cutoff are from the most massive high-mass star-forming regions in Cyg X: DR21, DR21(OH), W75N, DR15, and S106. Several possible scenarios can account for this discrepancy: (1) selection effect of existing interferometric observations since they are focused on high-mass star-forming clumps but not on the entire GMCs; (2) spacing filtering effect of interferometric observations that makes the lower-mass cores more difficult to detect; and (3) the high-mass star formation in Cyg X is periodic and is now in an inactive phase.

\subsection{Mass functions on larger scales}\label{subsec:larger}

Does such resemblance exist beyond the core scales? Studies using CO low-$J$ rotational transitions found that the mass functions of clouds on at least pc-scales deviate from the shape of the IMF by their shallower power-law tails \citep{1987ApJ...319..730S,1997ApJ...476..166W,2010ApJ...723..492R}. However, this may result from the observational biases from the different gas tracers \citep[e.g. CO is less sensitive to high-density regions compared with dust,][]{2008ApJ...684..395H}. To reveal the dependence of mass functions on the spatial scales of cloud structures, we degrade the resolution of the $N_{\rm H_2}$ map (Fig. \ref{fig:panorama}) by a factor of 2, 3, 4, 8, and 16, respectively, then generate the catalogs of cloud structures on each resolution with \emph{getsources} and derive their mass functions following the same analysis routine as above. Resultant statistical properties of these cloud structures and of their mass functions are summarized in Table \ref{tab:large}, and the mass functions of the cloud structures on different spatial scales are shown in Fig. \ref{fig:CMF_all_scale}. It is interesting to see that all the mass functions have similar shapes to the IMF, i.e., presenting a power-law high-mass part and a flattening in the low-mass regime, although the statistical significance drops for the largest spatial scales. This result indicates that the resemblance between the IMF and the mass functions of cloud structures (at least in the high-mass part) is not unique at core scales, but can be extended to at least several pc scales, which is not consistent with the idea that the shape of the IMF is directly inherited from the CMF.

\subsection{Origin of the CMF-IMF resemblance}\label{subsec:origin}

\begin{figure*}[htb!]
\epsscale{1}\plotone{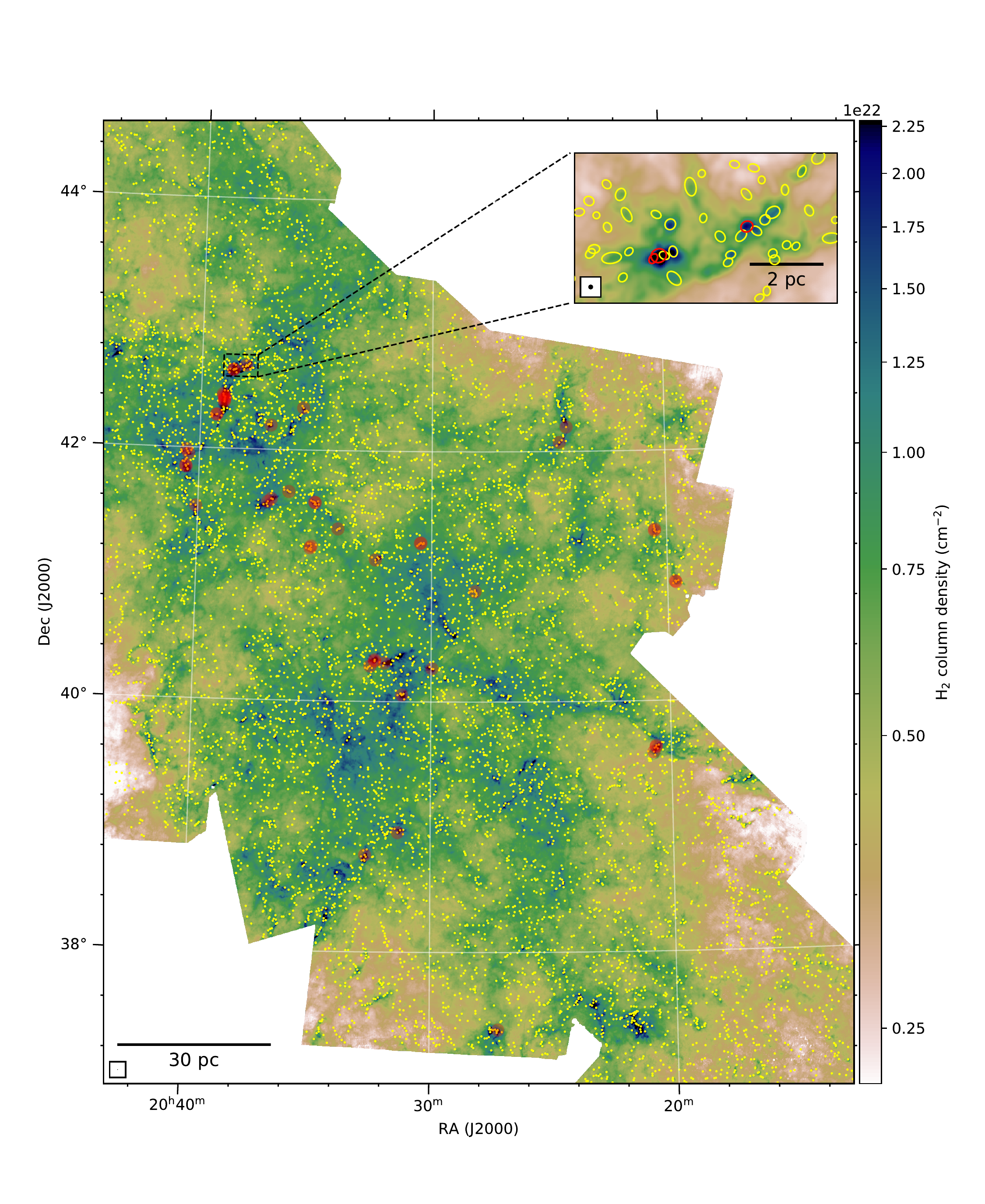}
\caption{A sample of \NumClumpGs\ cores in Cyg X overlaid on the $N_{\rm H_2}$ map from which they were extracted. In the main panel, positions of all the cores are marked as yellow dots and those as the targets of the SMA high-resolution survey (see Sect. \ref{subsec:origin}) are highlighted in red. The map derived from the \emph{Herschel} continuum images covers the whole giant molecular cloud with a dynamic range of 1,300 in spatial scale and 1,000 in column density, making it feasible to generate large samples of cloud structures. The zoom-in panel shows the FWHM ellipses of the cores with the same color coding as in the main panel. Resolution of the map (\beamColden) is shown in the lower-left corners of both the panels.}\label{fig:panorama}
\end{figure*}

\begin{figure*}[htb!]
\epsscale{.7}\plotone{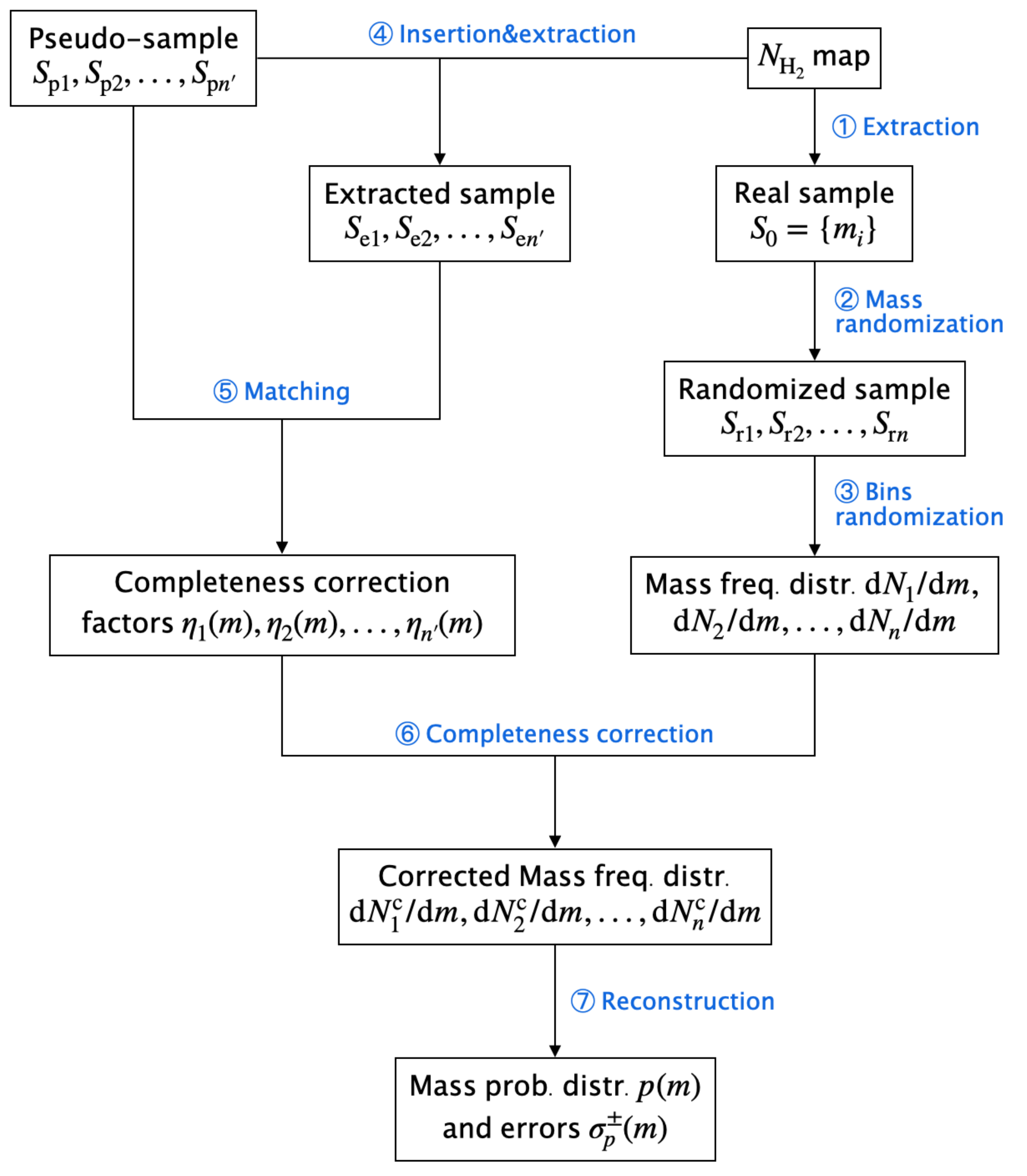}
\caption{Our designed procedure for deriving a statistically unbiased, ``true'' mass probability distribution and its uncertainties from a source sample $S_0$. Texts adhered to the arrow lines describe the steps of the procedure and those in black rectangles represent the intermediate and final results. See Sect. \ref{sec:analysis} for detailed descriptions.}\label{fig:flowchart}
\end{figure*}

\begin{figure*}[htb!]
\epsscale{.9}\plotone{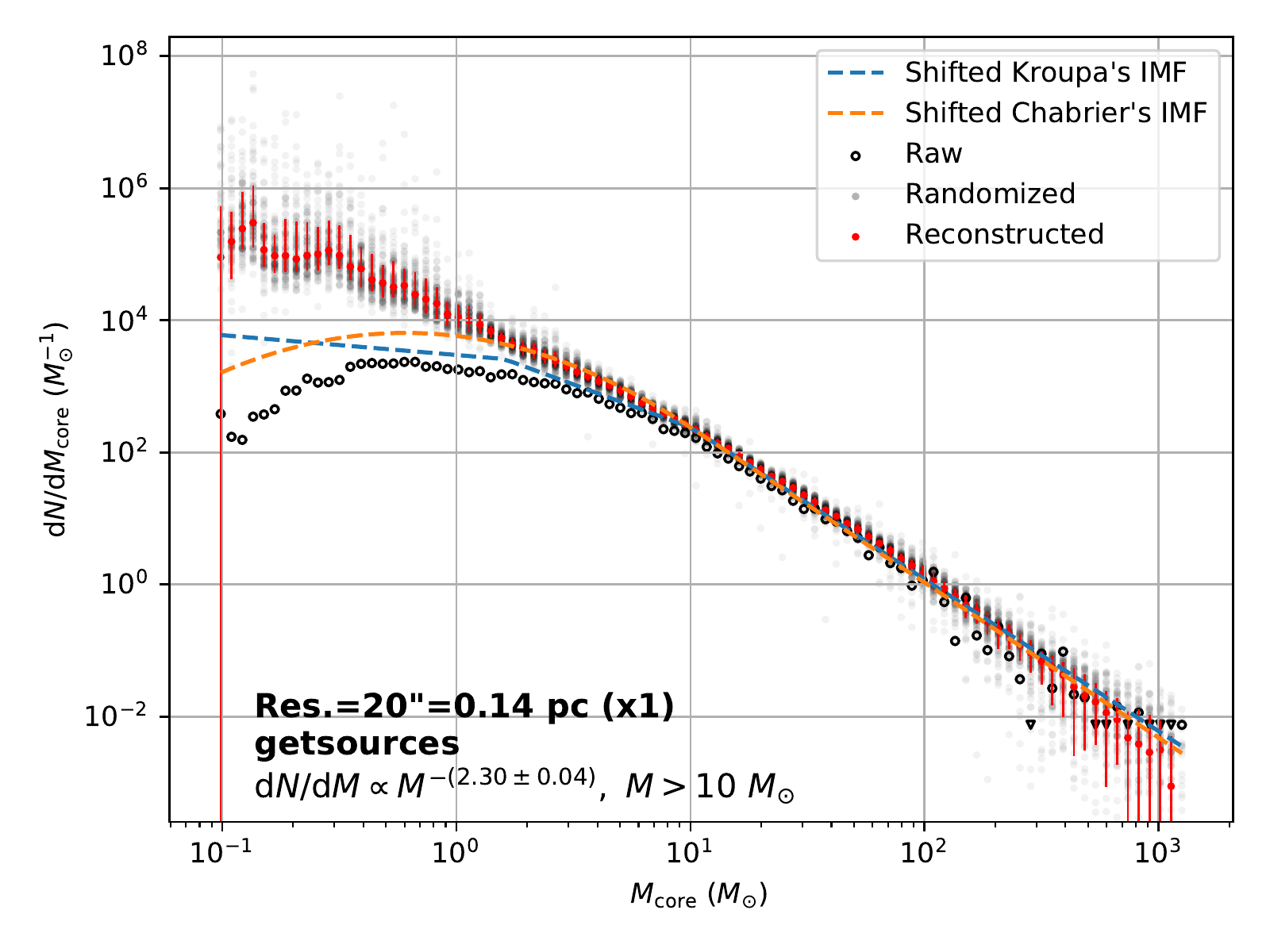}
\caption{The CMF (true mass probability distribution) of Cyg X derived with the procedure in Fig. \ref{fig:flowchart}. Red solid line and the associated shaded area illustrate the power-law slope and its error derived in the fitting procedure (see Sect. \ref{subsec:results}). Blue and orange dashed lines present the shifted IMFs of \citet{2001MNRAS.322..231K} and \citet{2005ASSL..327...41C}, respectively, with their starting points of the (most steep) power-law parts pinned at that of the CMF (10 \msun). The CMF has been corrected by the completeness levels, and its error bars are derived by considering the effects of mass uncertainty, binning uncertainty, and the uncertainty of completeness levels (see Sects. \ref{subsec:sample} and \ref{subsec:completeness}).} \label{fig:CMF_gs}
\end{figure*}

\begin{figure*}[htb!]
\epsscale{.9}\plotone{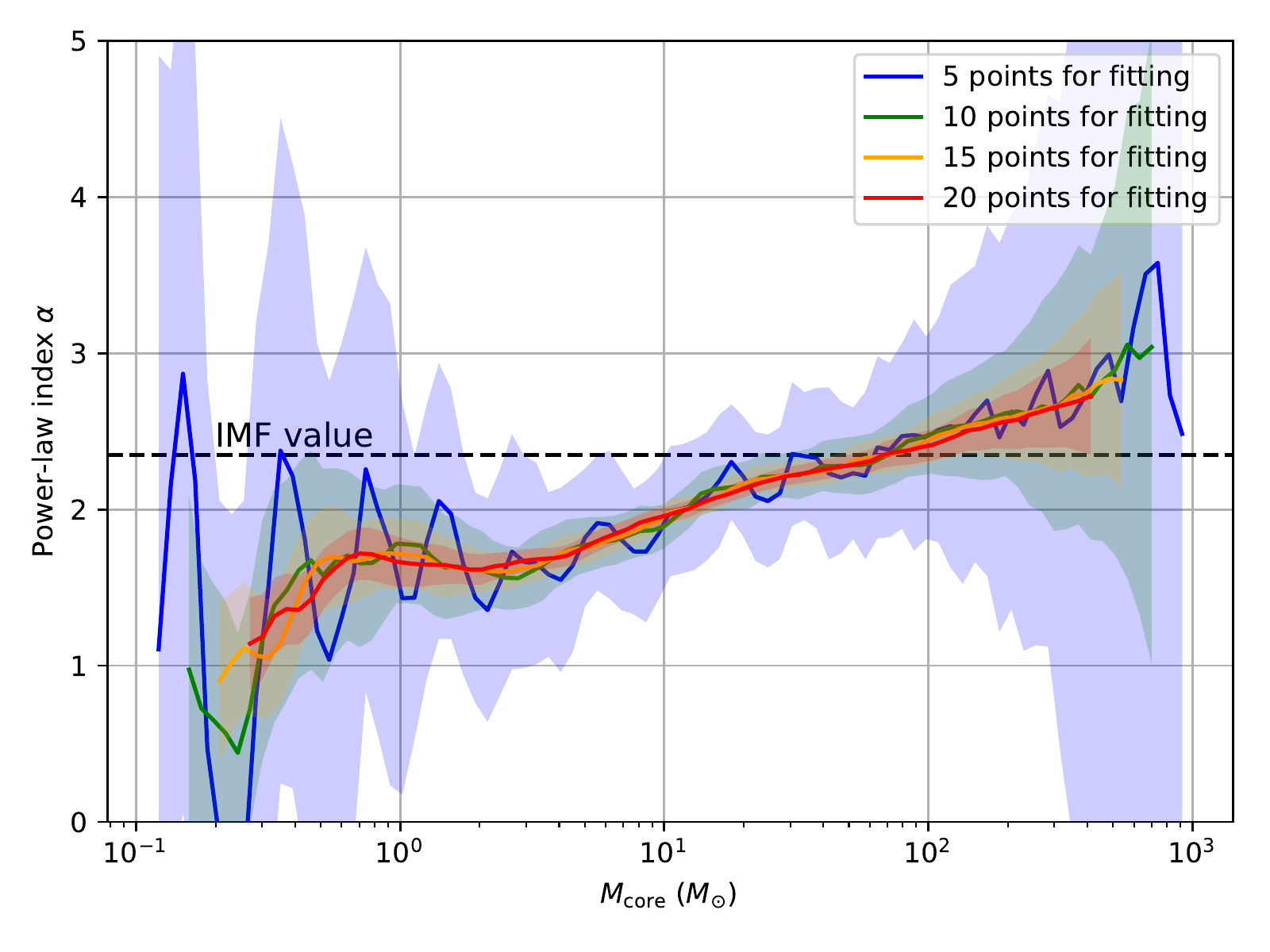}
\caption{Power-law index of the CMF in Fig. \ref{fig:CMF_gs} as a function of mass. Lines and shaded areas of different colors illustrate the index values and errors derived with different numbers of consecutive data points used for the fitting. Power-law index of the IMF (2.35) is shown as a horizontal dashed line.}\label{fig:slope}
\end{figure*}

How do cores distribute their masses into stars in light of the resemblance among their mass functions? Since it is impossible to trace the whole lifetime of a core observationally, here we focus on the observable core-to-condensation fragmentation to address this problem. Mathematically, there are infinite numbers of ways of core-to-condensation mass mapping even if the two mass functions are identical in shape (see Appendix \ref{app:infinite} for the proofs). Among all the possibilities two intuitive mapping scenarios have been proposed in the literature: (1) self-similar mapping \citep{1998A&A...336..150M,2007A&A...462L..17A,2018NatAs...2..478M}, where the probability that a core of mass $M_{\rm core}$ forms a condensation of mass $M_{\rm cd}$ is constant as long as $M_{\rm core}/M_{\rm cd}$ is constant; and (2) internal-IMF mapping \citep{2004Sci...303.1167B,2011MNRAS.410..788S}, where the underlying condensation mass function (CdMF) in each core is identical to the IMF. To check the validity of these scenarios, we conducted a high-resolution pilot survey toward \NumClumpGsCore\ selected cores with the Submillimeter Array (SMA) in 1.3 mm (see Appendix \ref{app:sma}), aimed at resolving the cores into condensations with a synthesized beam of $\sim$1\arcsec.8\ (0.012 pc@1.4 kpc), which is more than one order of magnitude smaller than the average core diameter. The target cores were chosen in the high-mass range of the core sample ($>$\mTransitGs\ \msun) for better detections and to better demonstrate the mass mapping in the more featured and reliable power-law part of the CMF. They were also selected to be absent from the strong free-free emissions of the ultracompact \hii\ regions in Cyg X \citep{2019ApJS..241....1C} so that the condensation masses can be properly estimated with the continuum fluxes which are dominated by thermal dust emissions. Fig. \ref{fig:panorama} presents the spatial distribution of the target cores in Cyg X and their 1.3-mm continuum images are shown in Fig. \ref{fig:sma_obs}. We use \emph{getsources} to identify condensations in the SMA images and selected \NumCore\ robust condensations with $\geq$5-$\sigma$ detections (see  Appendix \ref{app:sma}). To generate the core-condensation correspondence, we associated each condensation with the nearest core of which the FWHM ellipse covers the condensation. As a result, \NumCoreClump\ condensations are associated with the \NumClumpGsCore\ target cores. The FWHM diameters of these condensations fitted with 2-dimensional Gaussians range from \CoreSizeMin\ to \CoreSizeMax\ pc (cf. \ClumpGsSizeMin--\ClumpGsSizeMax\ pc for the cores). Using the dust temperature map of Cyg X in  \citep{2019ApJS..241....1C} and an approximate radiative transfer model \citep{2001A&A...365..440M,2018NatAs...2..478M}, we estimated the dust temperatures and masses of the condensations (see  Appendix \ref{app:sma}), which range in \CoreTMin--\CoreTMax\ K and \CoreMassMin--\CoreMassMax\ \msun, respectively. 

\begin{figure*}
\epsscale{1.2}\plotone{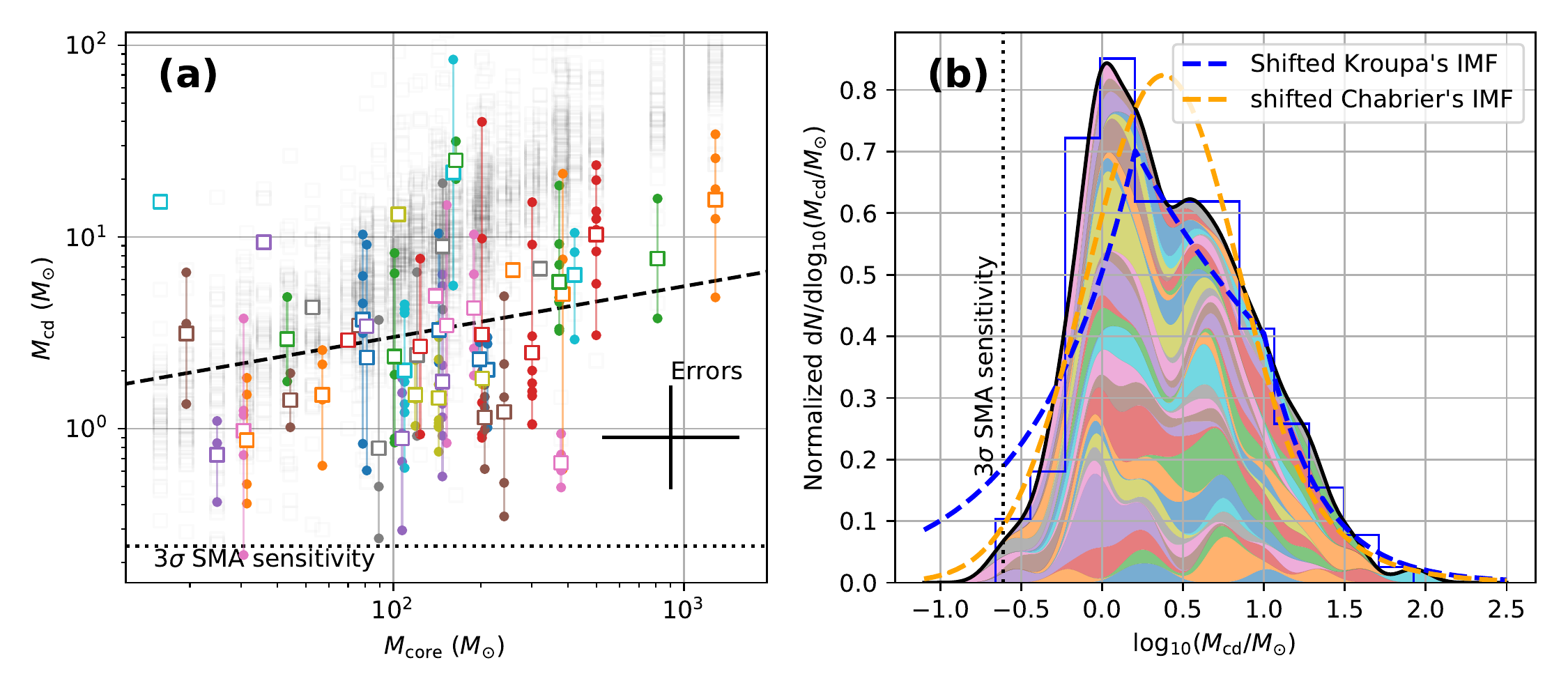}
\caption{Mass mapping from cores to condensations showing deviations from the predicted mapping scenarios. (a) Relation of core mass versus condensation mass. Each group of monochromatic dots associated by a vertical line represents the condensation masses in each core, with a square denoting their geometric mean value. An isolated square is shown if a core contains only one condensation. Regression analysis of the squares yields $M_{\rm cd, av}\propto M_{\rm core}^{\indexClumpMassCoreMass\pm\indexClumpMassCoreMassErr}$ (black dashed line), which shows different trend from the pseudo data of the self-similar mapping model generated with Monte-Carlo simulations (gray translucent squares). Averaged 3-$\sigma$ detection limit of the SMA maps derived with the noise levels and the parameters in the mass derivation (Appendix \ref{app:robust}) is shown as dotted lines. Typical error bars in standard deviation values are shown. (b) CdMF (solid black curve and also the blue histogram) of all the \NumCoreClump\ condensations and the CdMFs of individual cores as \emph{stacked} colored patches with the same color coding as in Panel (a) that contribute to the overall CdMF. CdMFs are generated by a kernel density estimation with a kernel width of 0.2 dex.}\label{fig:clump2core} 
\end{figure*}

Fig. \ref{fig:clump2core} presents the core mass versus condensation mass relation (Fig. \ref{fig:clump2core}a) and the contribution of individual CdMF of each core to the overall CdMF (Fig. \ref{fig:clump2core}b). Regression analysis on the mean condensation mass in each core versus the core mass yields $M_{\rm cd, av}\propto M_{\rm core}^{\indexClumpMassCoreMass\pm\indexClumpMassCoreMassErr}$ with a low correlation coefficient of $R=$\rValue, which is different from the $M_{\rm cd, av}\propto M_{\rm core}$ relation predicted by the self-similar mapping model (see  Appendix \ref{app:self-similar}). To further corroborate this difference, we ran Monte-Carlo simulations to generate pseudo condensation masses for each core following the self-similar mapping routine, with the condensation numbers and core-to-condensation formation efficiency mimicking the real data (see also Fig. \ref{fig:clump2core}). After 500 iterations we derive a mass relation with the pseudo data: $M_{\rm cd, av}\propto M_{\rm core}^{\indexClumpMassCoreMassMC\pm\indexClumpMassCoreMassMCErr}$, which indicates that the observed core-to-condensation mass mapping is highly unlikely to be reproduced by the self-similar model. To examine the validity of the internal-IMF model, we performed the two-sided Kolmogorov-Smirnov test with the core/condensation masses as input and with the null hypothesis that the CdMF in each core has the shape of the IMF regardless of the core mass. The derived p-value is \pValue, which indicates that this model can not reproduce the observed results. Moreover, in Fig. \ref{fig:clump2core}b, the mass distribution of condensations in each core greatly varies from case to case, and is clearly different from the IMF, apparently inconsistent with the internal-IMF model. The cumulative CdMF of all the condensations from all the observed cores has a shape relatively more comparable with the IMF, but there is still deviation in between. The evidence against both scenarios proposed in the literature points to the chaotic nature of the core-to-condensation (and probably also core-to-star) mass mapping process, and implies that any intuitive scenario proposed without full understanding of the underlying physics may oversimplify the reality and fail to explain it.

\section{Summary}\label{sec:conclusion}

With an unprecedentedly large sample generated in the Cyg X molecular cloud complex and our dedicated procedure of deriving the CMF and its uncertainty, we accurately reveal the shape of the CMF. We find that the CMF have a power-law tail with a slope index of $\slopeGs\pm\slopeErrorGs$, very close to that of the IMF values, while there is no significant flattening in the low-mass part as presented in IMFs. More detailed analyses illustrate that the slope of the CMF steepens with increasing masses, in contrast with the top-heavy IMFs discovered with recent interferometric observations. This can be explained by the incompleteness of the latter observations, space filtering effect of interferometers, or that the high-mass star formation of Cyg X is in its final stage. We also find that the similarity to the IMF is not unique for cores, but can be extended to pc-scale structures, indicating that this is a scale-free phenomenon. Our SMA observations further reveal that self-similar mass mapping may not be the case of the IMF origin from CMFs.


\acknowledgments 
\emph{Acknowledgments.}
Y.C., K.Q., Y.W. acknowledge National Natural Science Foundation of China (grant Nos. U1731237, 11629302, 11590781), and National Key R\&D Program of China No. 2017YFA0402600. Y.X. acknowledges the Fundamental Research Funds for the Central Universities 14380031, and National Natural Science Foundation of China grant 11101208. Y.C. is partially supported by the Scholarship No. 201906190105 of the China Scholarship Council and the Predoctoral Program of the Smithsonian Astrophysical Observatory (SAO).

\facilities{\emph{Herschel} (PACS, SPIRE), SMA (ASIC)}
\software{MIRIAD \citep{1995ASPC...77..433S}, getsources \citep{2012A&A...542A..81M}, clumpfind \citep{1994ApJ...428..693W}, Astropy \citep{2013A&A...558A..33A}, SciPy \citep{scipy}, Starlink \citep{2014ASPC..485..391C}, PyCupid \citep{2007ASPC..376..425B}}

\bibliographystyle{aasjournal}
\bibliography{ms}{}

\clearpage

\begin{deluxetable}{l|c|c|c|c|c|c|c}
\tabletypesize{\footnotesize}
\tablecaption{Samples of cloud structures on different spatial scales.}
\tablehead{\colhead{Source sample\tablenotemark{a}}& \colhead{gs-20\arcsec}& \colhead{gs-40\arcsec}& \colhead{gs-60\arcsec}& \colhead{gs-80\arcsec}& \colhead{gs-160\arcsec}& \colhead{gs-320\arcsec}& \colhead{cf-20\arcsec}}
\startdata
Robust source number& 8431& 2825& 1427& 821& 213& 42& 4479\\Median FWHM diameter (pc)& 0.28& 0.51& 0.72& 0.94& 1.73& 3.29& 0.29\\Median mass (\msun)& 3.0& 14.4& 33.1& 67.2& 322& 1262& 12.3\\Power-law starting point (\msun)& 10& 14& 28& 70& 400& 1000& 38\\Power-law index $\alpha$& 2.30$\pm$0.04& 2.23$\pm$0.05& 2.10$\pm$0.07& 2.16$\pm$0.11& 2.30$\pm$0.33& 1.89$\pm$0.70& 2.33$\pm$0.05\\
\enddata
\tablenotetext{a}{``gs'' and ``cf'' for samples derived with \emph{getsources} and \emph{clumpfind}, respectively. The numbers in arcseconds denote the resolutions of the maps from which the sources are extracted.}\label{tab:large}
\end{deluxetable}

\clearpage
\appendix 


\counterwithin{figure}{section}

\section{Details on deriving the true mass probability distributions and their uncertainties}\label{app:robust} 

\subsection{Mass uncertainty and randomization}\label{subapp:mass}

\begin{figure*}
\epsscale{.7}\plotone{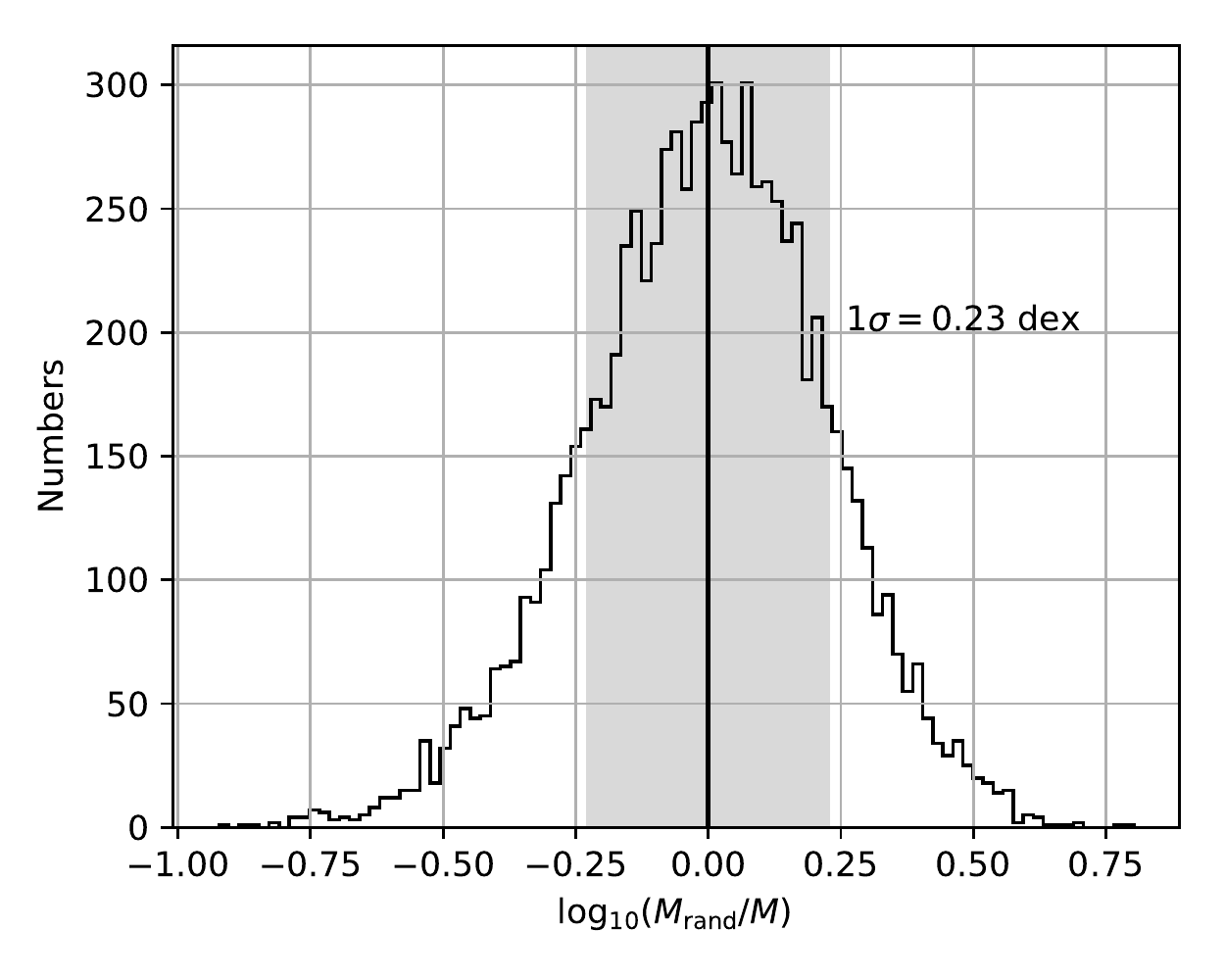}
\caption{Distribution of the mass randomization factors (i.e. random mass over real mass) with 8,000 Monte-Carlo realizations (see Appendix \ref{subapp:mass}). The shaded area shows the standard deviation (0.23 dex) of the distribution.}\label{fig:err_mass} 
\end{figure*}

Source masses derived by the extraction algorithms (\emph{getsources} or \emph{clumpfind}) can be equivalently derived with a graybody dust continuum model \citep{1983QJRAS..24..267H}

\begin{equation}
M=F_{\nu}D^2\Gamma/(\kappa_{\nu}B_{\nu}(T)), 
\end{equation}\label{eq:mass}

\noindent where $F_{\nu}$ is the flux density of the source at frequency $\nu$, $D=1400$ pc is the distance to the Sun, $\Gamma=100$ is the gas-to-dust mass ratio \citep{1983QJRAS..24..267H}, $B_{\nu}(T)$ is the Planck function, $T$ is the dust temperature, and $\kappa_{\nu}=\kappa_0(\nu/\nu_0)^\beta$ is the dust mass opacity \citep{1983QJRAS..24..267H}, which is evaluated following the HOBYS consortium\footnote{The \emph{Herschel} imaging survey of OB young stellar objects. \url{http://www.herschel.fr/cea/hobys/en/index.php}} \citep[e.g.][]{2017A&A...602A..77T}: $\kappa_0=10\ \rm cm^2g^{-1}$, $\nu_0=1$ THz, and the dust emissivity spectral index $\beta=2$. The uncertainty in mass mainly comes from the uncertainties in parameters $F_{\nu}$, $D$, $\Gamma$, $T$, and $\beta$. The 1-$\sigma$ uncertainties of $F_{\nu}$ and $T$ are adopted from  \citep{2019ApJS..241....1C} as $\sigma_{F_{\nu}}/F_{\nu}=$10\% and $\sigma_T=$2 K, respectively. The uncertainty of $D$ is evaluated as $\sigma_D=$200 pc based on the parallax measurement results in \citep{2012A&A...539A..79R}. The uncertainties of $\beta$ and $\Gamma$ are estimated as $\sigma_{\beta}=0.3$ and $\sigma_{\Gamma}=20$, respectively. To determine the uncertainty in mass estimates (see Sect. \ref{subsec:sample}), we evaluate these parameters as lognormally distributed random values with the standard deviations mentioned above, and calculate the mass randomization factors (i.e. the ratio between the simulated mass and the real mass) with Eq. \ref{eq:mass}. Fig. \ref{fig:err_mass} illustrates the distribution of the mass randomization factors after $\sim$8,000 Monte-Carlo realizations, which can be approximated by a lognormal distribution with a standard deviation of $\sim$0.23 dex, i.e. the mass uncertainty of our mass derivation is $\sim$0.23 dex ($\sim$55\%). The contributions of this mass uncertainty to the uncertainties of the mass probability distributions in Fig. \ref{fig:CMF_all_scale} are illustrated in Fig. \ref{fig:error_bar} as red curves (see also Sect. \ref{subsec:completeness}).

\subsection{Estimating completeness levels and their uncertainties}\label{subapp:complete}

For each of the seven source samples presented in Fig. \ref{fig:CMF_all_scale}, we perform 10 pseudo-source insertion-extraction experiments to estimate the completeness levels as a function of mass. In each experiment, 3000, 1000, 600, 400, 150, and 60 pseudo-sources modeled as 2D Gaussian functions are randomly positioned on the corresponding original/smoothed $N_{\rm H_2}$ maps on the $\times1$, $\times2$, $\times3$, $\times4$, $\times8$, and $\times16$ scales, respectively. The pseudo-sources have log-uniformly distributed masses covering the mass range of the real sources, and sizes following the distributions of the real source sizes. \emph{Getsources} and \emph{clumpfind} are used to extract the pseudo-sources on the $\times1$-16 maps and the $\times1$ map, respectively. Similar criteria as described in Sect. \ref{subsec:sample} are used to select robust sources in each experiment. See Table \ref{tab:simulation} for the detailed information of the source samples in all the experiments.   

\begin{figure*}[htb!]
\epsscale{1.1}\plotone{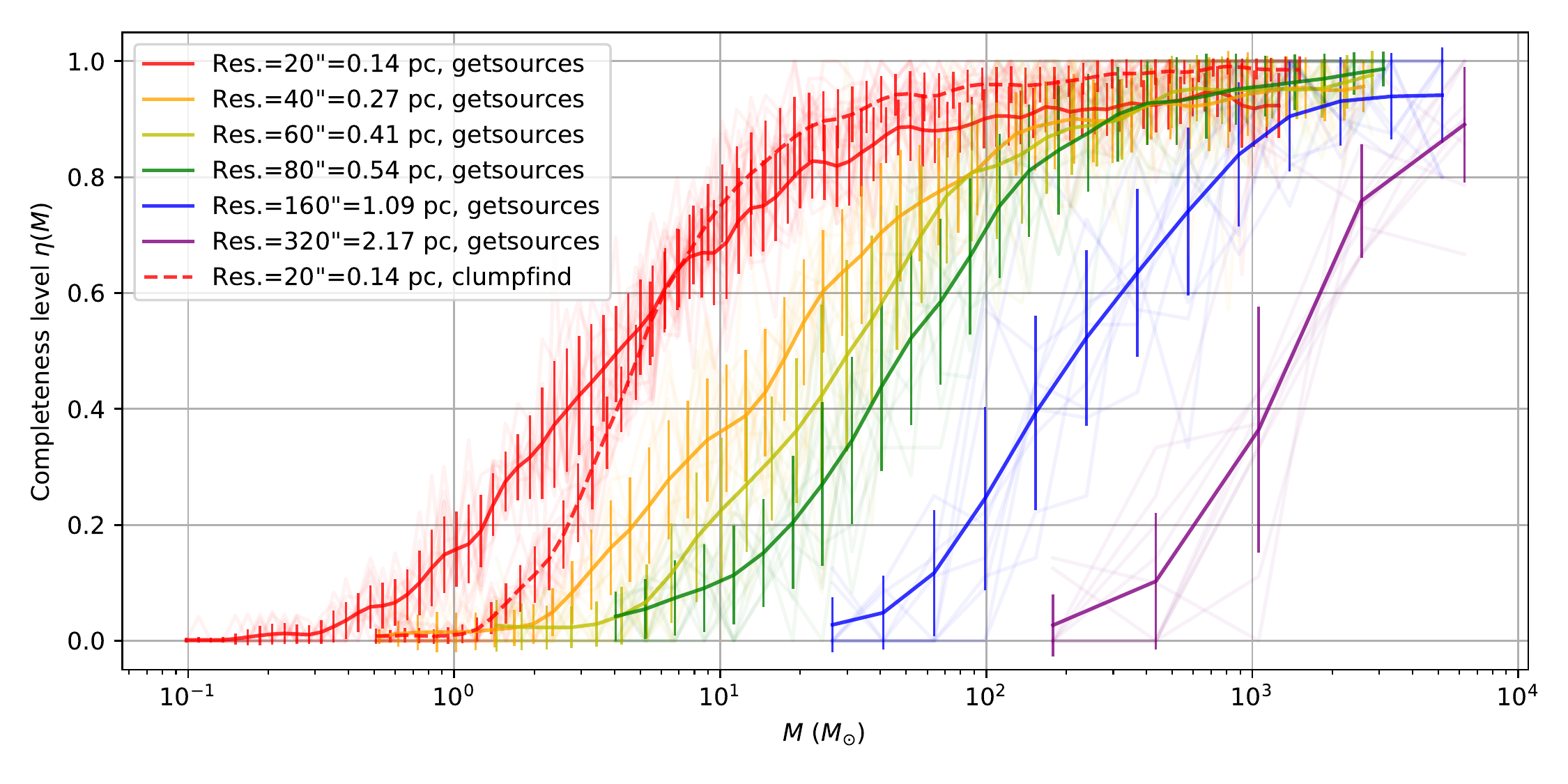}
\caption{Completeness levels of the samples of cloud structures as a function of mass, derived via pseudo-source insertion-extraction experiments (see Sect. \ref{subsec:completeness} and Appendix \ref{subapp:complete}). The pale-colored curves are completeness levels derived in individual experiments and the bright-colored ones are the means of the former. Error bars of the completeness levels are estimated as the standard derivations of the pale-colored curves. Only the pale-colored curves are used in the completeness correction process to include the contribution of the uncertainties of the completeness levels to the uncertainties of the derived mass probability distributions (see Sect. \ref{subsec:completeness}).}\label{fig:detect_rate}
\end{figure*}

\begin{figure*}[htb!]
\epsscale{1}\plotone{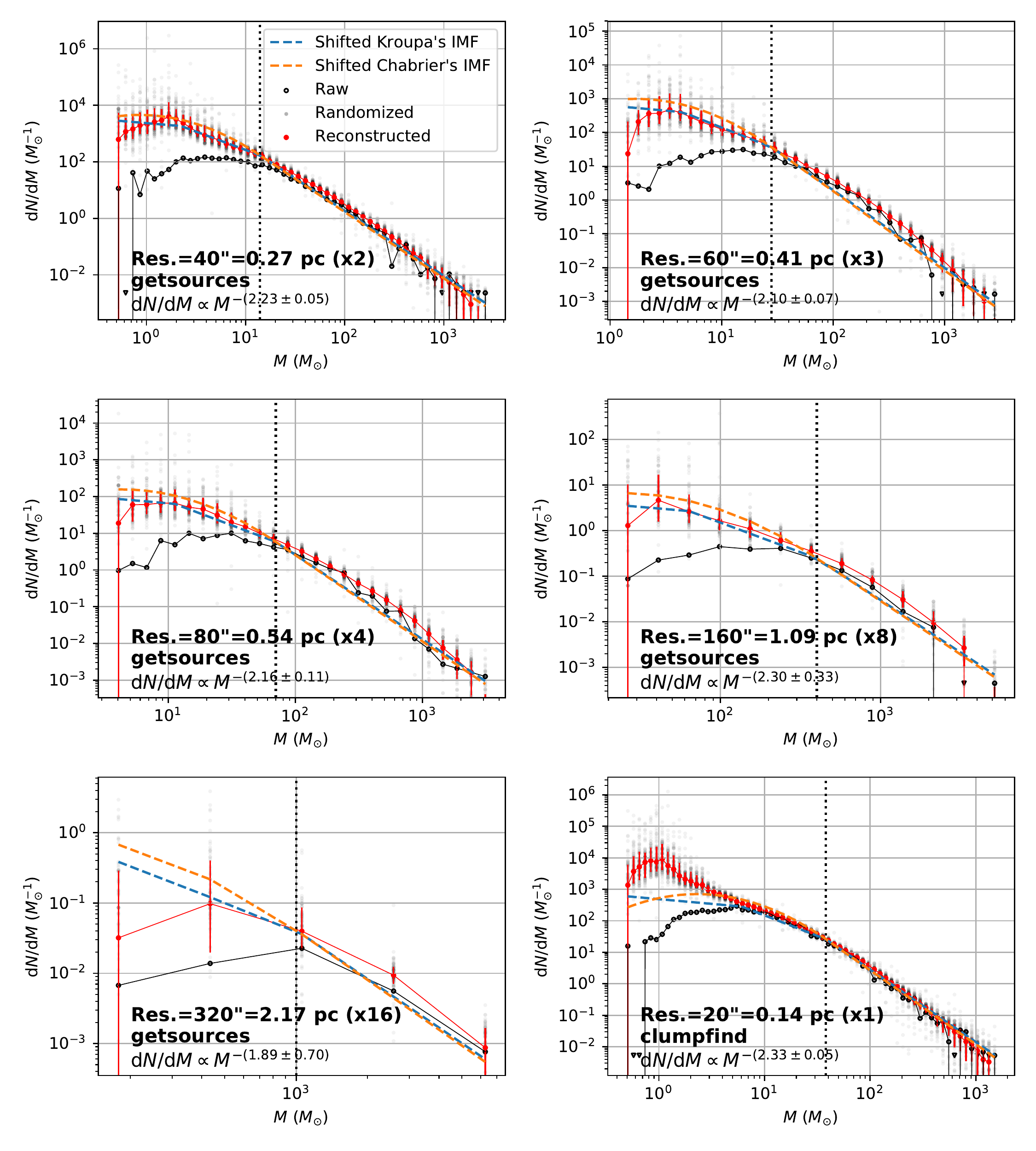}
\caption{Mass distributions of the samples of cloud structures. Raw distributions (black circles/triangles) are drawn using the source masses of the samples, with triangular symbols denoting zero values. Black transparent dots show the 100 randomized, completeness-corrected distributions of each sample (see Sect. \ref{sec:analysis}). Red dots and error bars present the reconstructed mass probability distributions and their uncertainties derived as the means and the standard deviations of the randomized distributions, respectively (see Sect. \ref{subsec:completeness}). Vertical dashed lines show the starting points of the power-law fitting of the high-mass parts and the fitting results are shown in each panel (see Sect. \ref{subsec:results}). Note that the mass probability distributions (red dots) are not normalized for better comparison.}\label{fig:CMF_all_scale}
\end{figure*}

\begin{figure*}
\epsscale{1}\plotone{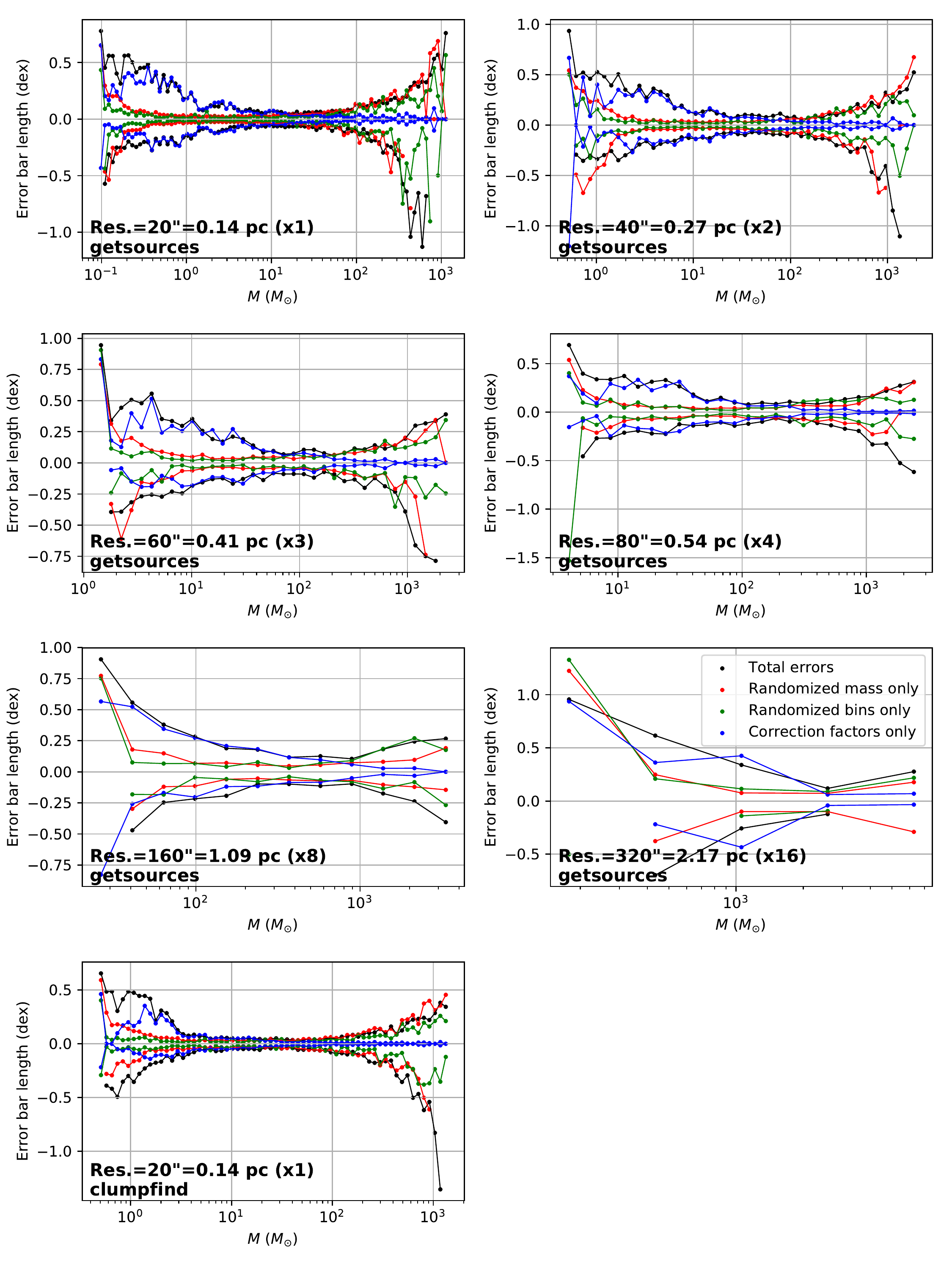}
\caption{Error bar lengths of the mass probability distributions in Fig. \ref{fig:CMF_all_scale} as a function of mass. Black, red, green, and blue curves represent the total errors, errors from mass uncertainties, errors from binning uncertainties, and errors from completeness levels, respectively (see Sect. \ref{subsec:completeness}). Note that the total errors are not derived from the other errors directly so they can happen to be smaller than the other ones in some plots.}\label{fig:error_bar} 
\end{figure*}

The extracted sources in an experiment contain both real and pseudo-sources. We spatially match the inserted pseudo-sources and the extracted sources with maximum allowed separations listed in Table \ref{tab:simulation}, with which the two matched sources can be considered as one based on eye inspection. The completeness level as a function of mass can then be calculated as the number fraction of the detected pseudo-sources with masses in a mass bin over the inserted pseudo-sources in that mass bin. Fig. \ref{fig:detect_rate} presents the completeness levels derived in all the experiments.

\section{High-resolution pilot survey resolving the cores into condensations}\label{app:sma}

High-resolution observations toward \NumClumpGsCore\ cores in the high-mass range ($M_{\rm core}\ge\mTransitGs$ \msun) of the core sample were conducted with the SMA at the 1.3 mm continuum band in the ``compact'' and ``extended'' array configurations during years 2015--2016 (project code: 2015A-S068, 2015A-S072, and 2016A-S061; PI: Keping Qiu). The target cores are absent from the strong free-free emissions from the ionized components of the ultracompact \hii\ regions in Cyg X \citep{2019ApJS..241....1C}, allowing us to better constrain the masses of the condensations with the 1.3 mm emissions. We used the Miriad \citep{1995ASPC...77..433S} software to conduct the data reduction and combine the data in the two configurations to make images. The final products contain 31 single-field images and two mosaic images (see Fig. \ref{fig:sma_obs}). These images recover the continuum emissions on spatial scales from their synthesized beams of $\sim$1\arcsec.8 (0.012 pc@1.4 kpc) to the largest angular scale of $\sim$20\arcsec\ (0.14 pc), the latter of which is very close to the resolution of the $N_{\rm H_2}$ map. The 1-$\sigma$ root-mean-square noise level of these images is $\sim$1.0 $\rm mJy\cdot beam^{-1}$ on average, evaluated over the emission-free regions. 

A blind extraction of \emph{getsources} on the 33 images yielded \NumCoreRaw\ detections, among which a large proportion are false detections caused by image noises and the artifacts on the image borders. With a criterion that source flux (before the primary-beam correction) must be above the 5-$\sigma$ values of the image noise levels, the bad detections can be easily removed and the sample is reduced to \NumCore\ robust condensations. Primary-beam correction was then applied to the fluxes and peak intensities of these condensations to eliminate the effect of the antenna angular response.  These condensations have FWHM diameters ranging from \CoreSizeMin--\CoreSizeMax\ pc, and 1.3 mm continuum fluxes ranging from \CoreFluxMin--\CoreFluxMax\ Jy. 

Dust temperatures of the condensations were estimated by combining the two results of (1) the large-scale (resolution$\sim$\beamT) dust temperature map of Cyg X in  \citep{2019ApJS..241....1C} (see their Fig. 26), and (2) an approximate radiative transfer model \citep{2001A&A...365..440M} for calculating the temperature distribution on small ($\le$0.1 pc) scales, following the idea of  \citep{2018NatAs...2..478M}. The former was produced together with the $N_{\rm H_2}$ map by fitting pixel-by-pixel the \emph{Herschel} continuum images of Cyg X, and the latter formulates a power-law radial temperature profile $T(r)\propto r^{-0.4}L_{\star}^{0.2}$ (see Eq. 2 of  \citep{2001A&A...365..440M}), where $L_{\star}$ is the luminosity of the central heating source (protostars). This decreasing temperature profile with radius will reach the large-scale temperature level at some radius $r_{\rm isoth}$ (see Eq. 3 of  \citep{2001A&A...365..440M}), out of which the temperature has negligible difference with the values in the large-scale temperature map. The actual temperature profile of a condensation is thus a piecewise function containing a power-law part for $r\le r_{\rm isoth}$ and a flat part for $r>r_{\rm isoth}$. If a condensation does not have associated protostars, $r_{\rm isoth}=0$ and the temperature profile contains only the flat part. The mean dust temperature of a condensation is estimated by averaging the temperature profile over the condensation volume. 

To determine $L_{\star}$ of the central protostars of the condensations, we spatially matched the condensation sample with the AllWISE Source Catalog \citep{2013yCat.2328....0C} and the \emph{Spitzer} Cygnus X Archive Catalog \citep{2010AAS...21541401K}. The AllWISE catalog was used first due to its better spatial coverage and fewer photometrically saturated sources, and the \emph{Spitzer} catalog was used as supplementary data in case of missing infrared sources. With a maximum allowed separation of 0.01 pc, \NumCoreIR\ out of \NumCore\ condensations are associated with the AllWISE/\emph{Spitzer} infrared sources. We integrated the fluxes over all the \emph{WISE}/\emph{Spitzer} bands for the \NumCoreIR\ condensations to estimate their $L_{\star}$, which ranges from \CoreLIRMin\ to \CoreLIRMax\ \lsun. As a result, the derived temperatures of the \NumCore\ condensations range from \CoreTMin\ to \CoreTMax\ K, with a median value of \CoreTMed\ K. 

With the condensation temperature determined, condensation mass can be derived with the same dust emission model as in the derivation of the core mass: $M_{\rm cd}=F_{\nu}D^2\Gamma/(\kappa_{\nu}B_{\nu}(T))$, where $\nu=230$ GHz. The other parameters in this equation took the same evaluation as in the derivation of core masses. The resultant masses of the \NumCore\ condensations range from \CoreMassMin\ to \CoreMassMax\ \msun, with a median value of \CoreMassMed\ \msun. Similar to the method of determining the core mass uncertainty, we conducted Monte-Carlo simulations with the following evaluation of the parameter errors: SMA flux uncertainty $\sigma_{F_{\nu}}/F_{\nu}=$10\%, $\sigma_D=200$ pc, $\sigma_{\beta}=0.3$, $\sigma_T=4$ K, and $\sigma_{\Gamma}=20$, and the condensation mass uncertainty is estimated to be $\sim$0.26 dex after 500 simulation runs. 

\section{Infinite possible ways of self-similar mass mapping}\label{app:infinite}

Here we illustrate that there are infinite possible ways of core-to-condensation mass mapping given that CdMF is identical to CMF. Let $F(m_{\rm cl})$ and $G(m_{\rm c})$ be the normalized CMF and CdMF, respectively, and $p(m_{\rm cl},m_{\rm c}){\rm d}m_{\rm c}$ be the probability that a core of mass $m_{\rm cl}$ forms condensations in the mass interval [$m_{\rm c}$, $m_{\rm c}+{\rm d}m_{\rm c}$]. We have $\int F(m_{\rm cl}){\rm d}m_{\rm cl}=\int G(m_{\rm c}){\rm d}m_{\rm c}=\int p(m_{\rm cl},m_{\rm c}){\rm d}m_{\rm c}=1$ due to normalization. By definition $G(m_{\rm c})$ can be written as

\begin{equation}\label{eq:cmf}
G(m_{\rm c})=\int F(m_{\rm cl})p(m_{\rm cl},m_{\rm c}){\rm d}m_{\rm cl}.
\end{equation}

\noindent Assuming that CdMF is identical to CMF, i.e., $F=G$, and that $p(m_{\rm cl},m_{\rm c})$ can be written as

\begin{equation}
p(m_{\rm cl},m_{\rm c})=G(m_{\rm c})f(m_{\rm cl}), 
\end{equation}

\noindent then Eq. \ref{eq:cmf} becomes

\begin{equation}
1=\int F(m_{\rm cl})f(m_{\rm cl}){\rm d}m_{\rm cl}.
\end{equation}

Since $F(m_{\rm cl})$ is normalized, there are infinite possible ways of choosing $f(m_{\rm cl})$ to satisfy the above equation (c.f. there are infinite possible ways of choosing $a_i$ to satisfy $\sum a_ix_i=1$ given $\sum x_i=1$). In other words, there are infinite ways of mass mapping even if the two mass functions are identical.

\section{Properties of the self-similar mass mapping}\label{app:self-similar}

Here we illustrate that for self-similar core-to-condensation mass mapping, the mean condensation mass of a core is proportional to the core mass. Self-similarity requires that the probability that a core of mass $m_{\rm cl}$ forms condensations of mass $m_{\rm c}$ is equal to the one that a core of mass $tm_{\rm cl}$ forms condensations of mass $tm_{\rm c}$, i.e. 

\begin{equation}
p(m_{\rm cl},m_{\rm c}){\rm d}m_{\rm c}=p(tm_{\rm cl},tm_{\rm c}){\rm d}tm_{\rm c},
\end{equation}

\noindent where $p(m_{\rm cl},m_{\rm c})$ the probability density, and $t$ is the mass scaling factor. That is,

\begin{equation}
p(m_{\rm cl},m_{\rm c})=tp(tm_{\rm cl},tm_{\rm c}).
\end{equation}

\noindent By definition, $p(m_{\rm cl},m_{\rm c})$ is normalized for all $m_{\rm cl}$, i.e. $\int p(m_{\rm cl},m_{\rm c}){\rm d}m_{\rm c}=1$. For the core of mass $m_{\rm cl}$, the mean mass of condensations that it produces is

\begin{equation}
\overline{m}_{\rm c}=\int m_{\rm c}p(m_{\rm cl},m_{\rm c}){\rm d}m_{\rm c}.
\end{equation}

\noindent For the core of mass $tm_{\rm cl}$ the mean condensation mass is

\begin{equation}
\overline{m}_{\rm c}'=\int tm_{\rm c}p(tm_{\rm cl},tm_{\rm c}){\rm d}tm_{\rm c}=\int tm_{\rm c}\frac{p(m_{\rm cl},m_{\rm c})}{t}{\rm d}tm_{\rm c}=t\int m_{\rm c}p(m_{\rm cl},m_{\rm c}){\rm d}m_{\rm c}=t\overline{m}_{\rm c},
\end{equation}

\noindent which means that $\overline{m}_{\rm c}$ scales as $m_{\rm cl}$ with the same factor $t$, i.e., $\overline{m}_{\rm c}$ is proportional to $m_{\rm cl}$.


\clearpage

\startlongtable
\begin{deluxetable}{c|c|ccc|ccc}
\tabletypesize{\footnotesize}
\tablecaption{Generation and extraction of the samples of pseudo-sources.}
\tablehead{
    \colhead{Sample\tablenotemark{a}}& 
    \colhead{Pseudo-source No.\tablenotemark{b}}& 
    \colhead{Min mass\tablenotemark{c}}& 
    \colhead{Border width\tablenotemark{d}}& 
    \colhead{Extracted source No.\tablenotemark{e}}& 
    \colhead{Max separation\tablenotemark{f}}& 
    \colhead{Matched source No.\tablenotemark{g}}& 
    \colhead{Matching rate\tablenotemark{h}}\\
    \colhead{}& 
    \colhead{}& 
    \colhead{(\msun)}& 
    \colhead{(\arcsec)}& 
    \colhead{}& 
    \colhead{(pc)}& 
    \colhead{}& 
    \colhead{}}
\startdata
gs-20\arcsec-1& 3000& 0.001& 22& 12570& 0.025& 1711& 57.0\%\\
gs-20\arcsec-2& 3000& 0.001& 22& 11981& 0.025& 1659& 55.3\%\\
gs-20\arcsec-3& 3000& 0.001& 22& 12060& 0.025& 1645& 54.8\%\\
gs-20\arcsec-4& 3000& 0.001& 22& 12042& 0.025& 1602& 53.4\%\\
gs-20\arcsec-5& 3000& 0.001& 22& 12180& 0.025& 1623& 54.1\%\\
gs-20\arcsec-6& 3000& 0.001& 22& 12258& 0.025& 1652& 55.1\%\\
gs-20\arcsec-7& 3000& 0.001& 22& 12032& 0.025& 1660& 55.3\%\\
gs-20\arcsec-8& 3000& 0.001& 22& 12084& 0.025& 1648& 54.9\%\\
gs-20\arcsec-9& 3000& 0.001& 22& 12116& 0.025& 1660& 55.3\%\\
gs-20\arcsec-10& 3000& 0.001& 22& 11769& 0.025& 1598& 53.3\%\\
\hline
gs-40\arcsec-1& 1000& 0.001& 44& 4218& 0.05& 531& 53.1\%\\
gs-40\arcsec-2& 1000& 0.001& 44& 4032& 0.05& 546& 54.6\%\\
gs-40\arcsec-3& 1000& 0.001& 44& 3973& 0.05& 534& 53.4\%\\
gs-40\arcsec-4& 1000& 0.001& 44& 4081& 0.05& 556& 55.6\%\\
gs-40\arcsec-5& 1000& 0.001& 44& 4076& 0.05& 544& 54.4\%\\
gs-40\arcsec-6& 1000& 0.001& 44& 4020& 0.05& 551& 55.1\%\\
gs-40\arcsec-7& 1000& 0.001& 44& 4059& 0.05& 538& 53.8\%\\
gs-40\arcsec-8& 1000& 0.001& 44& 4125& 0.05& 537& 53.7\%\\
gs-40\arcsec-9& 1000& 0.001& 44& 4088& 0.05& 525& 52.5\%\\
gs-40\arcsec-10& 1000& 0.001& 44& 4026& 0.05& 540& 54.0\%\\
\hline
gs-60\arcsec-1& 600& 0.01& 66& 2243& 0.1& 331& 55.2\%\\
gs-60\arcsec-2& 600& 0.01& 66& 2219& 0.1& 320& 53.3\%\\
gs-60\arcsec-3& 600& 0.01& 66& 2262& 0.1& 335& 55.8\%\\
gs-60\arcsec-4& 600& 0.01& 66& 2188& 0.1& 344& 57.3\%\\
gs-60\arcsec-5& 600& 0.01& 66& 2220& 0.1& 332& 55.3\%\\
gs-60\arcsec-6& 600& 0.01& 66& 2254& 0.1& 331& 55.2\%\\
gs-60\arcsec-7& 600& 0.01& 66& 2241& 0.1& 325& 54.2\%\\
gs-60\arcsec-8& 600& 0.01& 66& 2283& 0.1& 348& 58.0\%\\
gs-60\arcsec-9& 600& 0.01& 66& 2242& 0.1& 340& 56.7\%\\
gs-60\arcsec-10& 600& 0.01& 66& 2186& 0.1& 323& 53.8\%\\
\hline
gs-80\arcsec-1& 400& 0.01& 88& 1475& 0.2& 241& 60.2\%\\
gs-80\arcsec-2& 400& 0.01& 88& 1462& 0.2& 221& 55.2\%\\
gs-80\arcsec-3& 400& 0.01& 88& 1492& 0.2& 222& 55.5\%\\
gs-80\arcsec-4& 400& 0.01& 88& 1482& 0.2& 221& 55.2\%\\
gs-80\arcsec-5& 400& 0.01& 88& 1469& 0.2& 248& 62.0\%\\
gs-80\arcsec-6& 400& 0.01& 88& 1450& 0.2& 233& 58.2\%\\
gs-80\arcsec-7& 400& 0.01& 88& 1501& 0.2& 224& 56.0\%\\
gs-80\arcsec-8& 400& 0.01& 88& 1445& 0.2& 232& 58.0\%\\
gs-80\arcsec-9& 400& 0.01& 88& 1454& 0.2& 238& 59.5\%\\
gs-80\arcsec-10& 400& 0.01& 88& 1453& 0.2& 218& 54.5\%\\
\hline
gs-160\arcsec-1& 150& 0.1& 177& 475& 0.4& 81& 54.0\%\\
gs-160\arcsec-2& 150& 0.1& 177& 485& 0.4& 79& 52.7\%\\
gs-160\arcsec-3& 150& 0.1& 177& 484& 0.4& 76& 50.7\%\\
gs-160\arcsec-4& 150& 0.1& 177& 480& 0.4& 81& 54.0\%\\
gs-160\arcsec-5& 150& 0.1& 177& 501& 0.4& 88& 58.7\%\\
gs-160\arcsec-6& 150& 0.1& 177& 499& 0.4& 81& 54.0\%\\
gs-160\arcsec-7& 150& 0.1& 177& 473& 0.4& 72& 48.0\%\\
gs-160\arcsec-8& 150& 0.1& 177& 487& 0.4& 76& 50.7\%\\
gs-160\arcsec-9& 150& 0.1& 177& 491& 0.4& 83& 55.3\%\\
gs-160\arcsec-10& 150& 0.1& 177& 497& 0.4& 90& 60.0\%\\
\hline
gs-320\arcsec-1& 60& 10& 354& 154& 0.8& 28& 46.7\%\\
gs-320\arcsec-2& 60& 10& 354& 169& 0.8& 32& 53.3\%\\
gs-320\arcsec-3& 60& 10& 354& 167& 0.8& 28& 46.7\%\\
gs-320\arcsec-4& 60& 10& 354& 152& 0.8& 28& 46.7\%\\
gs-320\arcsec-5& 60& 10& 354& 156& 0.8& 27& 45.0\%\\
gs-320\arcsec-6& 60& 10& 354& 157& 0.8& 29& 48.3\%\\
gs-320\arcsec-7& 60& 10& 354& 164& 0.8& 30& 50.0\%\\
gs-320\arcsec-8& 60& 10& 354& 157& 0.8& 24& 40.0\%\\
gs-320\arcsec-9& 60& 10& 354& 170& 0.8& 29& 48.3\%\\
gs-320\arcsec-10& 60& 10& 354& 168& 0.8& 25& 41.7\%\\
\hline
cf-20\arcsec-1& 3000& 0.001& 22& 5757& 0.025& 1723& 57.4\%\\
cf-20\arcsec-2& 3000& 0.001& 22& 5695& 0.025& 1647& 54.9\%\\
cf-20\arcsec-3& 3000& 0.001& 22& 5696& 0.025& 1628& 54.3\%\\
cf-20\arcsec-4& 3000& 0.001& 22& 5654& 0.025& 1617& 53.9\%\\
cf-20\arcsec-5& 3000& 0.001& 22& 5656& 0.025& 1626& 54.2\%\\
cf-20\arcsec-6& 3000& 0.001& 22& 5644& 0.025& 1631& 54.4\%\\
cf-20\arcsec-7& 3000& 0.001& 22& 5719& 0.025& 1663& 55.4\%\\
cf-20\arcsec-8& 3000& 0.001& 22& 5677& 0.025& 1652& 55.1\%\\
cf-20\arcsec-9& 3000& 0.001& 22& 5690& 0.025& 1687& 56.2\%\\
cf-20\arcsec-10& 3000& 0.001& 22& 5617& 0.025& 1593& 53.1\%\\
\hline

\enddata
\tablenotetext{a}{``gs'' and ``cf'' for samples derived with \emph{getsources} and \emph{clumpfind}, respectively. The first number in arcseconds denotes the resolution of the map from which the sources are extracted. The second number is the number of the source insertion-extraction experiment.}
\tablenotetext{b}{Number of pseudo-sources inserted in one experiment.}
\tablenotetext{c}{Minimum mass used to exclude sources with implausibly small extracted mass values.}
\tablenotetext{d}{Sources with distances from the map borders smaller than these values are excluded to avoid artifacts.}
\tablenotetext{e}{Number of extracted robust sources (including both pseudo-sources and real sources) from a map in an experiment.}
\tablenotetext{f}{Maximum separation in parsec (assuming a distance of 1.4 kpc) used to match the extracted sources and the pseudo-sources.}
\tablenotetext{g}{Number of matched pseudo-sources.}
\tablenotetext{h}{Equals to matched source No./pseudo-source No., which is an estimation of the detection rate of the pseudo-sources.}
\label{tab:simulation}
\end{deluxetable}

\clearpage

\begin{figure*}[htb!]
\epsscale{.9}\plotone{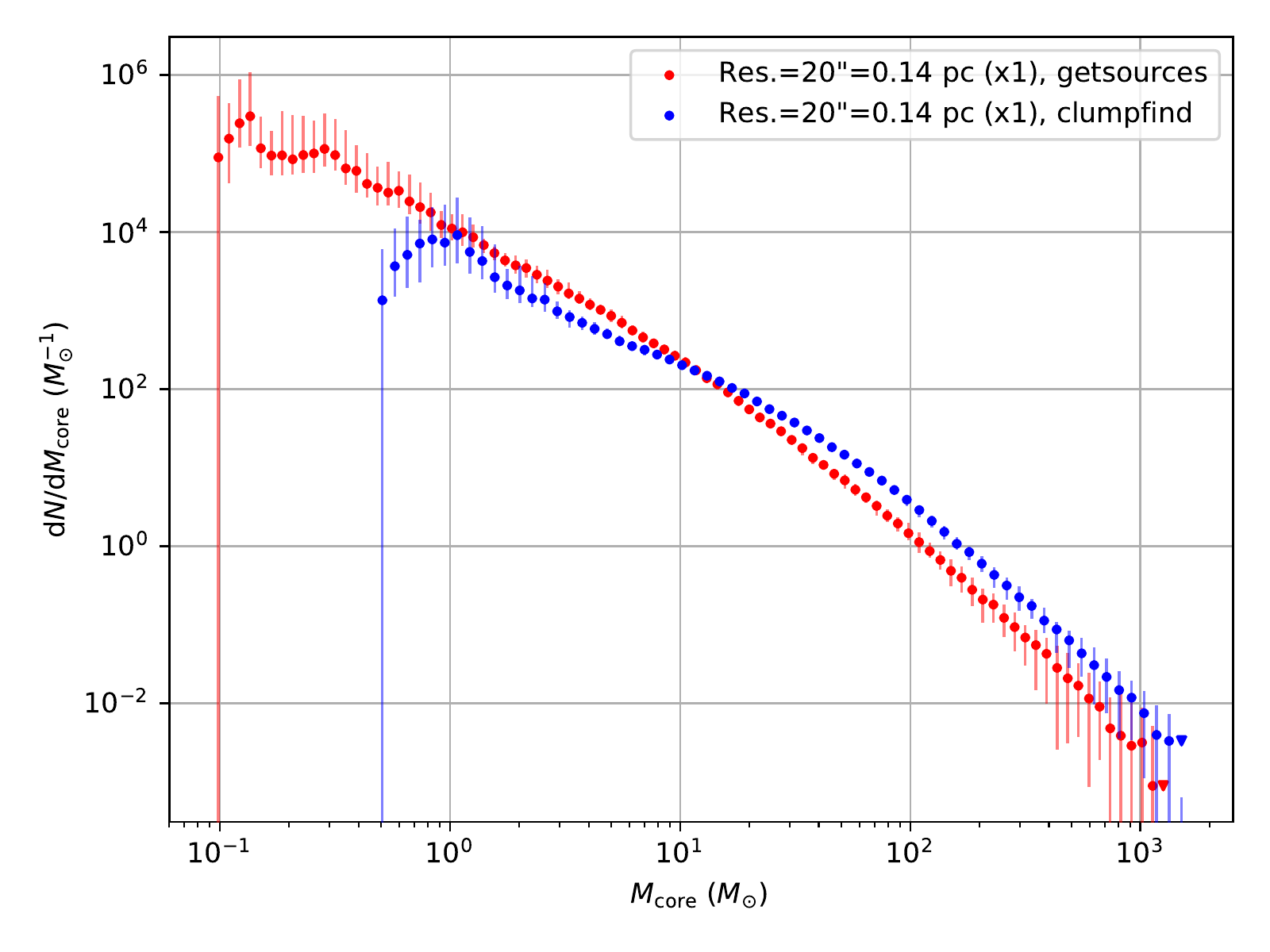}
\caption{Mass probability distributions of the cores derived with \emph{getsources} and \emph{clumpfind}, respectively, for comparison (see Sect. \ref{subsec:clumpfind}).}\label{fig:compare}
\end{figure*}

\begin{figure*}[htb!]
\epsscale{1}\plotone{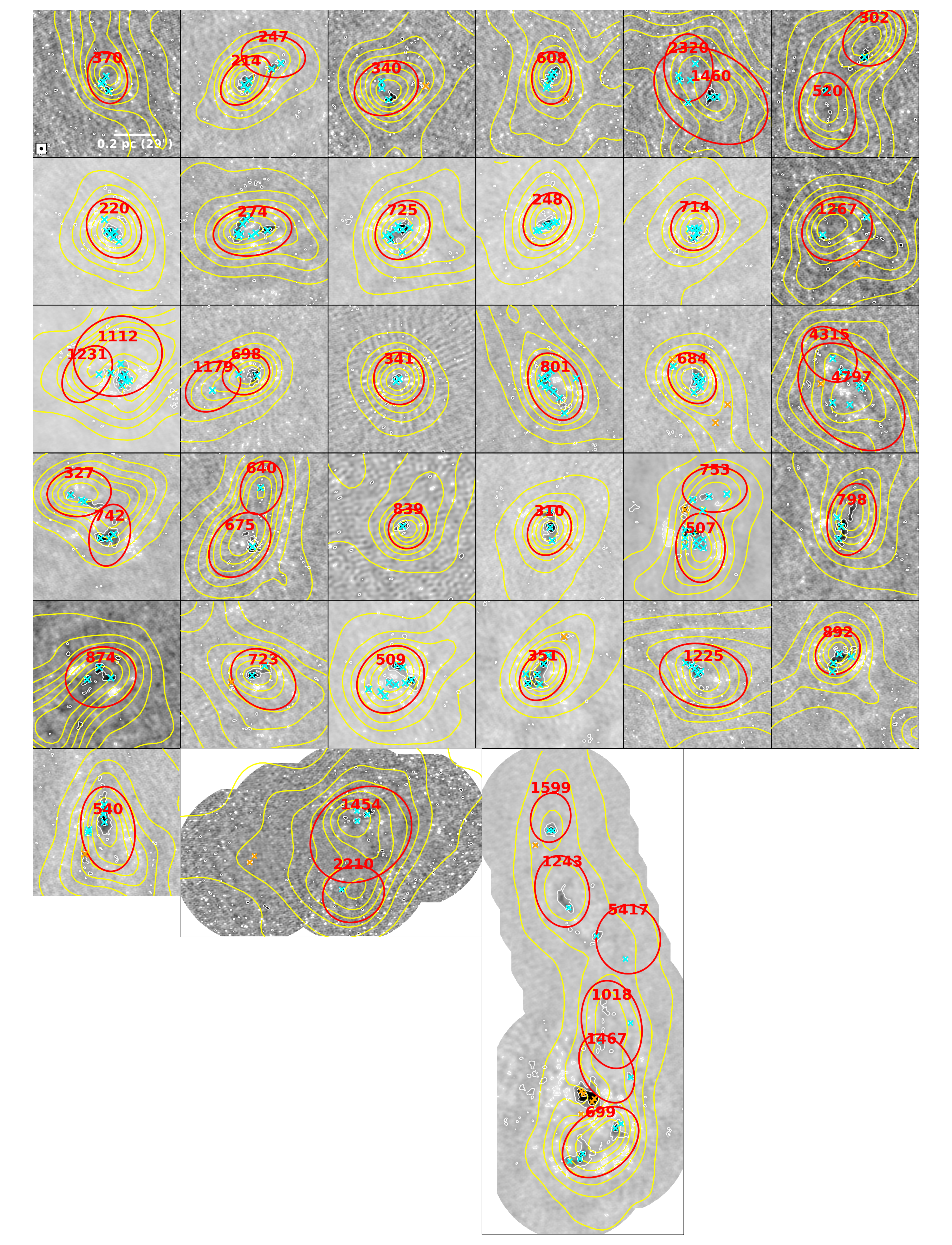}
\caption{High-resolution 1.3 mm continuum images of SMA observations showing the substructures of the \NumClumpGsCore\ target cores. Grayscales are the 1.3 mm continuum emissions, with higher intensities rendered darker. North is up and East is left. Red ellipses show the FWHM diameters of the cores. Cyan/orange crosses mark the condensations associated/not associated with the cores. Yellow contours of the $\rm H_2$ column density derived from the \emph{Herschel} data are drawn linearly from the minimum to the maximum values in the panels. White contours of the 1.3 mm emissions are drawn in $[..., -4^2, -4^1, -4^0, 4^0, 4^1, 4^2, ...]\times3\sigma$, with negative contours shown in white dashed curves. All the panels share the same scale bar and the angular resolution illustrated in the most top-left panel.}\label{fig:sma_obs}
\end{figure*}

\end{document}